# *High-g-Factor Phase-Matched Circular Dichroism of Second Harmonic Generation in Chiral Polar Liquids*

*Xiuhu Zhao[1], Jinxing Li[1], Mingjun Huang[1,2], Satoshi Aya\*[1,2]*

[1]South China Advanced Institute for Soft Matter Science and Technology (AISMST), School of Emergent Soft Matter, South China University of Technology, Guangzhou, 510640, P. R. China

[2]Guangdong Provincial Key Laboratory of Functional and Intelligent Hybrid Materials and Devices, South China University of Technology, Guangzhou, 510640, P. R. China

*Abstract*

Circular dichroism is a technologically important phenomenon contrasting the absorption and resultant emission properties between left- and right-handed circularly polarized light. While the chiral handedness of systems mainly determines the mechanism of the circular dichroism in linear optics, the counterpart in the nonlinear optical regime is nontrivial. Here, in contrast to traditional nonlinear circular dichroism responses from structured surfaces, we report on an unprecedented bulk-material-induced circular dichroism of second harmonic generation with a massive *g*-factor up to 1.8.

*Introduction*

Chirality is an ubiquitous property of the manifestation of asymmetry, existing in a wide range of scales in many science branches, such as matter structure and light.[1, 2] In light physics, the optical chirality or activity arises especially in matters without mirror symmetry (e.g., chiral systems), covering several branches of optical properties such as optical rotation (OR) and circular dichroism (CD), respectively. Due to their technological importance, these phenomena have attracted tremendous attention.[2-4] Especially, CD refers to the difference in the absorption between the left- and right-handed circularly polarized light, which arises either in linear or nonlinear optical regimes. In the linear optical regime, the chiral sense and magnitude of CD are governed by the atomic or molecular arrangement of functional groups with absorption in enantiomeric molecules, chiral biological structures, chiral helices, and asymmetric metasurfaces.[5-12] On the other hand, CD produced in the nonlinear optical regime is governed by nonlinear interactions between light and matter arising from the asymmetry of matter.[13-15] The nonlinear CD effects for many systems are much more effective than the linear counterparts, which benefit the generation of strong optical activity.[16-18] Also, the nonlinear CD offers an essential route for studying and visualizing biological tissues in high-contrast and low-artifacts.[19-22] Therefore, developing high-efficient nonlinear CD materials for high-intensity optoelectronic applications and microscopy measurements is highly demanding.

The CD effect of the second harmonic generation, dubbed SHG-CD, is one of the central targets in the exploration of nonlinear CD materials. Unlike the linear CD effect differentiating the absorptions of the left- and right-handed circularly polarized light, the SHG-CD effect generates a contrast in the magnitude of SH signals that are excited by left- and right-handed circularly polarized fundamental optical fields.[23-25] Therein, a unique combination of the frequency doubling property of SHG and the strong chirality-sensitive emission in SHG-CD would allow many potential applications like bio-imaging,[19-22] photoelectric sensors, high-contrast circular polarizers, and holographic displays.

In early times, it was discovered that chiral molecules (e.g., 2,2'-dihydroxy-1,1'-binaphthyl) adsorbed on the air/water interface typically result in SHG, which is sensitive to the handedness of the fundamental optical field.[16] Later, similar phenomena were found in many chiral systems [13, 26-28], biological collagens and tissues,[20, 22] and chiral organic-inorganic micro-nano structured materials.[29-32] Additionally, several achiral structures have shown the SHG-CD effect triggered by extrinsic chirality induced by oblique incident waves.[17, 18, 33, 34] Recall that most current SHG-CD phenomena are primarily based on electric and magnetic dipole interactions in non-flowing solid systems.[14, 23, 30, 32, 35, 36] It has rarely been explored whether the SHG-CD effect would be achieved through the nonlinear phase-matching (PM) conditions due to the limited material systems before.

The recent discovery of novel ferroelectric nematic ($N_F$) liquid crystals with high fluidity and outstanding nonlinear optical response in flowing liquid matters are significant.[37-44] The chiral counterpart of $N_F$, i.e., the helielectric nematic liquid crystals (HN*), exhibits chiral periodic polarization structures.[45, 46] Proper polarization periodicities in HN* state can significantly improve nonlinear optical conversion efficiency through nontrivial phase-matching mechanisms.[47] Here, we report that PM-enabled SHG amplification in the HN* is highly chirality sensitive, exhibiting considerably large SHG-CD values. We also explored the tunability of SHG-CD under an electric field. This unprecedented scenario is a generic strategy for all potential polarization systems (if polarization structures can be controlled), thus providing a route for developing flexible nonlinear optical technologies.

*Results and Discussion*

Figure 1 summarizes how the combinations of the handedness of fundamental circular polarization of light and the periodicity of HN* structures affect the output of SHG. When the medium processes a periodic rectangular structure with simple upward and downward polarizations, the quasi-phase-matching (QPM) condition occurs under the condition of $W = 2L_c = \lambda/2(n_\omega - n_{2\omega})$, where the periodicity of the periodic rectangular structure $W$ equals the twice of the optical coherence length $L_c$, $n_\omega$ and $n_{2\omega}$ are the refractive indices at the wavelengths of the fundamental and SH waves. For this case, the SH intensity is strong and increases with the medium thickness,[48] but does not exhibit SHG-CD. Once chirality is introduced, the helicity is coupled to the nematic polarity, leading to a continuous rotation of the polarization field. The chiral nature of the polarization helices allows the system to exhibit strong SHG-CD under proper conditions. Starting from the PM-satisfied structure, when the helical periodicity $p$ is reduced slightly by incorporating a chiral dopant into the system, the SHG itself dramatically decreases due to the deviation from the PM condition. SHG-CD is not recognizable in this case. As we further reduce the helical periodicity to about $L_c - 1.7L_c$, SHG enhances again due to the satisfied phase-matching of the HN* polarization helix.[47] Under this circumstance, the circular polarized fundamental optical field with the same handedness as that of the HN* polarization helix exhibits considerably stronger SHG than that for the opposite handedness of polarization. Namely, a gigantic SHG-CD effect emerges. Further decrease of the helical pitch leads to a complete mismatch of the phases between the fundamental and SH waves. Also, the reduction of the helical pitch near the sub-micron range ($p$ is close to or short than the wavelength of the fundamental wave) directly reduces the effective polarity of the system. As a result, the SH signal almost vanishes at the short limit of the helical pitch.

To experimentally address the SHG-CD of the HN* state, we prepared HN* consisting of a well-known $N_F$ compound, RM734, and chiral dopants (R811 and S811 for right- and left-handed systems, respectively).[45] For satisfying the phase-matching condition for the HN* structure, we doped 1.1wt% R811 or 1.1wt% S811 into RM734, where the helical pitch is about 6-8 μm (so $p \sim L_c - 1.7L_c$) in the HN* phase for both systems. We employed a parallel beam for incidence and used half-wave and quarter-wave plates to adjust the handedness of the circular polarization of the fundamental wave (see Figure S2 for the measurement optics). Figure 2a,b compares the temperature dependencies of the SH intensity for 1.1%R811/RM734 and 1.1%S811/RM734 at right- and left-handed circularly polarized light incidence, respectively. The LC cell thickness was fixed to be 50 μm. The upper and lower surfaces were rubbed syn-parallelly. The samples were slowly cooled down from the isotropic phase below a 1 K/min cooling rate, guaranteeing the strong polarization anchoring parallel to the rubbing direction. The helical axis is along the surface normal direction. Figure S1 shows the polarizing light microscopy (PLM) textures of 1.1% R811/RM734 and 1.1% S811/RM734 in both the high-temperature apolar N* (i.e., cholesteric) and low-temperature HN* phases, demonstrating a large-area homogeneous texture. While the apolar N* phase is SHG-inactive, entering the HN* phase gives rise to a continuous growth of the SH signal due to both the steady increase of polarization order and the helical pitch $p$ approaches to $L_c - 1.7L_c$ upon decreasing temperature (achieving nearly $p \sim L_c - 1.7L_c$ at about 100-130 °C). This consequence strongly suggests that when a linear polarized fundamental wave along the rubbing direction of the incident face is used for the SH excitation, there is no difference in SH signal magnitude between the right- and left-handed polarization helices. Both the 1.1% R811/RM734 and 1.1% S811/RM734 systems emit strong SH signals. However, when circular polarization is used for the SH excitation, we see a mirror relationship between the two systems. Namely, only when the handedness of the circular polarization of the fundamental wave is consistent with that of the HN* helix, strong SH signal is observed. The contrast between the SH signals for opposite handedness of the fundamental wave yields a high anisotropy factor $g_{SHG} = 2\frac{I^L_{2\omega} - I^R_{2\omega}}{I^L_{2\omega} + I^R_{2\omega}}$ which defines how much SHG is sensitive to the handedness of circularly polarized incoming light.[16,17] $I^L_{2\omega}$ and $I^R_{2\omega}$ stands for the SH intensities of the left- and right-handed circularly polarized fundamental waves, respectively. Then, we obtain $g$-factor as a function of temperature for both systems (Figure 2c). The maximum $g$-factor reaches 1.82 and -1.8 for 1.1% S811/RM734 and 1.1% R811/RM734, respectively. The opposite signs of the $g_{SHG}$ for the two systems mean the preferred chirality of the system for emitting a strong SH signal is reversed.

Figure 3a,b demonstrate the thickness dependencies of phase-matched SH signals for 1.1%R811/RM734 and 1.1%S811/RM734 under left- and right-handed circularly polarized light incidences at 110 °C (good condition for HN* phase-matching around $p \sim L_c - 1.7L_c$). We used syn-parallel-rubbed wedge-shaped cells, which enables us to continuously track the variation of the SH signal as a function of the sample thickness. The insets in Figure 3a,b show the PLM textures in the HN* phase in the wedge-shaped cells. The so-called Cano lines arrange periodically, which separates the neighboring domains with the variation of the number of the helix by a full helical pitch.[45] Since the number of helices is constant in each Cano domain, the actual helical pitch is distributed around the thermal equilibrium pitch value by changing the sample thickness along the wedge direction.[49,50] Overall, the textures are homogeneous over a millimeter scale with few defect lines crossing the Cano lines, indicating a good quality of the alignment of the HN* helices. Consistent with the observation in Figure 2, the SH intensity increases dramatically with thickness only when the handedness of the circular polarization of the fundamental wave is the same as that of the HN* helix through the chiral HN* phase-matching pathway. Notably, the SH intensity oscillates periodically around the average SH intensity curve (blue and pink solid lines). The oscillation periodicity as a variation of the sample thickness is about 5.5 μm, corresponding to the variation of the lateral distance, 1200 μm, along the wedge direction in the texture of the wedge-shaped cell. The lateral distance is consistent with the distance between two neighboring Cano lines, suggesting that the oscillation of the SH signal is due to the phase variation between the fundamental and SH waves caused by the change of the real HN* helical pitch in the wedge-shaped cell. The *g*-factor exhibits a clear dependence on the sample thickness: $g_{SHG}$ is very low and increases almost linearly with increasing the sample thickness up to about 6 μm (Figure S3); at larger sample thicknesses (>~6 μm), the averaged $|g_{SHG}|$ saturates at about 1.7-1.8 with a small signal oscillation aforementioned. Note that the HN* in the wedge confinement forms the first complete helix structure around 4-6 μm.

Figure 3c,d represents the direct visualization of the strong SHG-CD effect in HN* using a normal digital single-lens reflex camera (DSLR; Nikon D7000). The thickness of the LCs cell was 50 μm, and the temperature was kept at 110 °C. The exposure time of DSLR is short enough for 0.05 s, so clarifying that the chirality-sensitive SHG emission is even apparent to the naked eye. As for the SHG emission performance, the conversion efficiency of SHG in the 50 μm sample is about 0.007% at maximum. This enables us to estimate the corresponding conversion efficiency at 2 cm (typical length scale for nonlinear optical crystals) by assuming almost a linear SH signal growth with a sample thickness of about 3%. Of course, in real devices, the saturation of the SH signal is expected over a specific sample thickness due to scattering and absorption issues, which should be addressed in the future with the exploration of suitable device engineering methodology.

The linear coupling between the local polarization and the electric field in the HN* state gifts the system to exhibit tunability on the polarization structure and the resulting electro-optic properties. Figure 4 demonstrates the electric field tunability and reconfigurability of SHG. The electric-field switching of the SHG-CD signals in a 50-μm cell. The applied DC electric field parallels the surface polarization (Figure 4a). While the apolar N* state exhibits the dielectric response with symmetric structural variation to the electric field,[51] the HN* state takes an asymmetric structural change under the DC bias.[47, 52] The half-pitch areas with the averaged polarization parallel to the DC field tend to extend along the helical axis, and the rest with the averaged polarization antiparallel shrinks (Figure 4a). Figure 4b shows the SH intensities as a function of time during the electric field switched on. When an electric field of 1 mV/μm was applied, the SH intensity of both the HN* systems rapidly decreased, during which the *g*-factor also decreased (see Figure S4 for the variasiotn of the *g*-factor of 1.1%R811/RM734). This consequence is because the asymmetric HN* structure (as drawn in Figure 4a; bottom) induced by the in-plane DC electric field leads to the phase unmatching condition. When the electric field is removed, with the HN* structure returning to the initial state, the SH signals and *g*-factor recover under the satisfaction of phase-matching conditions the SH signals.

*Conclusion*

In summary, we use a nontrivial phase-matching technique to achieve a strong second harmonic circular dichroism in helielectric nematic liquid crystals. The HN* has a *g*-factor up to 1.8, and this strong circular dichroism has visualization. Furthermore, we investigated the effect of an applied electric field on circular dichroism, and our findings demonstrated that the SHG-CD of HN* is electrically tunable and reconfigurable. HN* materials, with their high mobility and ease of preparation, offer several practical utilities in the design of novel flexible nonlinear optical devices and chiral probes.

*Experimental Section*

*Materials*

All commercial chiral dopants and solvents were used as received. The $N_F$ LC RM734 and the chiral dopants used to prepare the HN* LCs are shown in Figure S5. RM734 was synthesized according to the method in Ref. [37]. We used two commercially-available chiral dopants, R811 (right-handed) and S811 (left-handed) (from Shanghai Acmec Biochemical Co., Ltd). The chiral dopants and RM734 were weighed at a mass ratio of 1.1:98.9, dissolved in the solvent chloroform, and dried in an oven at 55 °C for 24 h.

*LC cell preparation*

We used homemade LC cells for PLM texture observation and SHG measurements. KPI-3000 (Shenzhen Haihao Technology Co., Ltd) was used as the planar aligner. The HN* materials were sandwiched between two planarly-rubbed glass plates. If not otherwise stated, the rubbing directions of both the top and bottom surfaces are parallel to each other, i.e., syn-polar buffing. The LC cell thickness was adjusted in the range of 2-100 μm by using silica .

*Measurement of SHG properties*

The light source for inducing SHG is a pulsed laser (MPL-III-1064-20μJ, Changchun New Industries Optoelectronics Tech. Co., Ltd.; wavelength: 1064 nm, energy: 20 μJ, repetition rate: 100 Hz). We use a half-wave and quarter-wave plates to adjust the state of the incident light to

right- or left-handed circular polarization before entering to samples (Figure S2). A photomultiplier tube is used to collect the SH signal (DH-PMT-D100V, Daheng New Epoch Technology, Inc.). To measure the temperature dependence of the SH signal, we use a homemade LabVIEW program to control the temperature (temperature variation less than 0.1 °C), and record the SH signal by an oscilloscope at a 1 °C interval.

*Thickness dependence of SHG*

We used the same configuration as the SHG measurement system. In order to conduct the thickness dependence of the SHG experiment, the temperature is kept at 110 °C, a Piezo Inertia Stage and controller (PD1 and KIM001, Thorlabs Inc.) were used to control the movement of the LC cell along the thickness direction, so the SH intensity was scanned by changing the measuring position along the wedge direction. A homemade wedge-shaped liquid crystal cell with the thickest end of 100 μm was used for the thickness SHG measurement.

*SHG Visualization*

We used the same configuration as in the normal SHG measurement. A 50-μm thick planar LC cell was used for SHG visualization, and the temperature is kept at 110 °C. The transmitted SH light is collected by a digital single-lens reflex camera (D7000, Nikon Corporation). The exposure time is 0.05 s.

*Electric switching of SH intensity*

A 50-μm LC cell is used. Glass surfaces are etched with interdigitated comb-shaped ITO electrodes, and the electrode width and gap between adjacent electrodes are 1 mm. This enables us to apply an in-plane electric field perpendicular to the helical axis of the helielectrics. The two wires were connected to a function generator. A homemade LabVIEW program for voltage-controlled SHG measurement is conducted. The direct current electric field is set to 2 volts and the time interval of the pickup points is set to be 5000 ms. The temperature is kept at 110 °C.

*Acknowledgements*

S.A. and M.H. acknowledge the National Key Research and Development Program of China (No. 2022YFA1405000), the Recruitment Program of Guangdong (No. 2016ZT06C322), and the 111 Project (No. B18023). S.A. acknowledges the supports from the International Science and Technology Cooperation Program of Guangdong province (No. 2022A0505050006), General Program of Guangdong Natural Science Foundation (No. 2022A1515011026), Guangzhou Basic and Applied Basic Research Foundation (No. 202102021156) and the Fundamental Research Funds for the Central University (No. 2022ZYGXZR001). M.H.

acknowledges the support from the National Natural Science Foundation of China (NSFC No. 52273292).

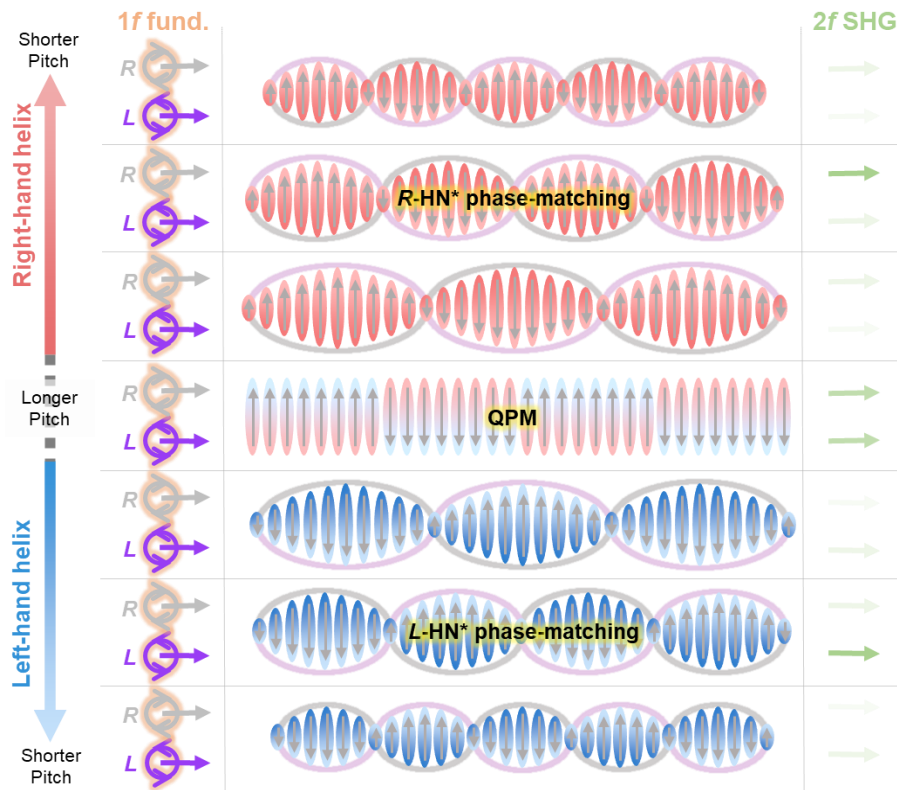

*Figure 1*. Schematics drawing how the circular polarizations of the fundamental light through the HN* structures with different helical pitches affect the output of the SH signals. The visibility of the green arrows at the exist of the materials indicates the magnitude of the SH signals. *L*-HN* and *R*-HN* phase-matching mean the phase matching of the HN* state occurs for left-handed and right-handed circular polarization of the fundamental wave, respectively. Green arrows indicate the magnitude of the output SH waves with the corresponding transparency for each condition.

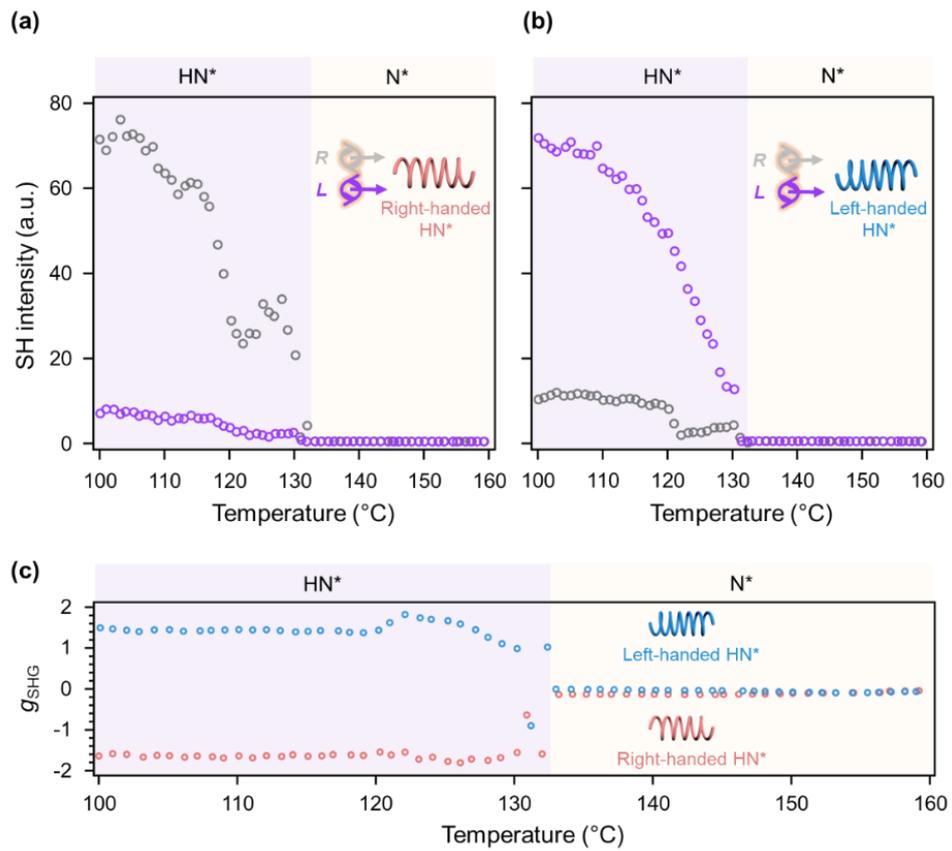

*Figure 2*. a) and b) The temperature dependencies of SH intensities of right-handedness HN* 1.1%R811/RM734 and left-handedness HN* 1.1%S811/RM734, respectively. c) The *g*-factor of right-handedness HN* 1.1%R811/RM734 and left-handedness HN* 1.1%S811/RM734, respectively.

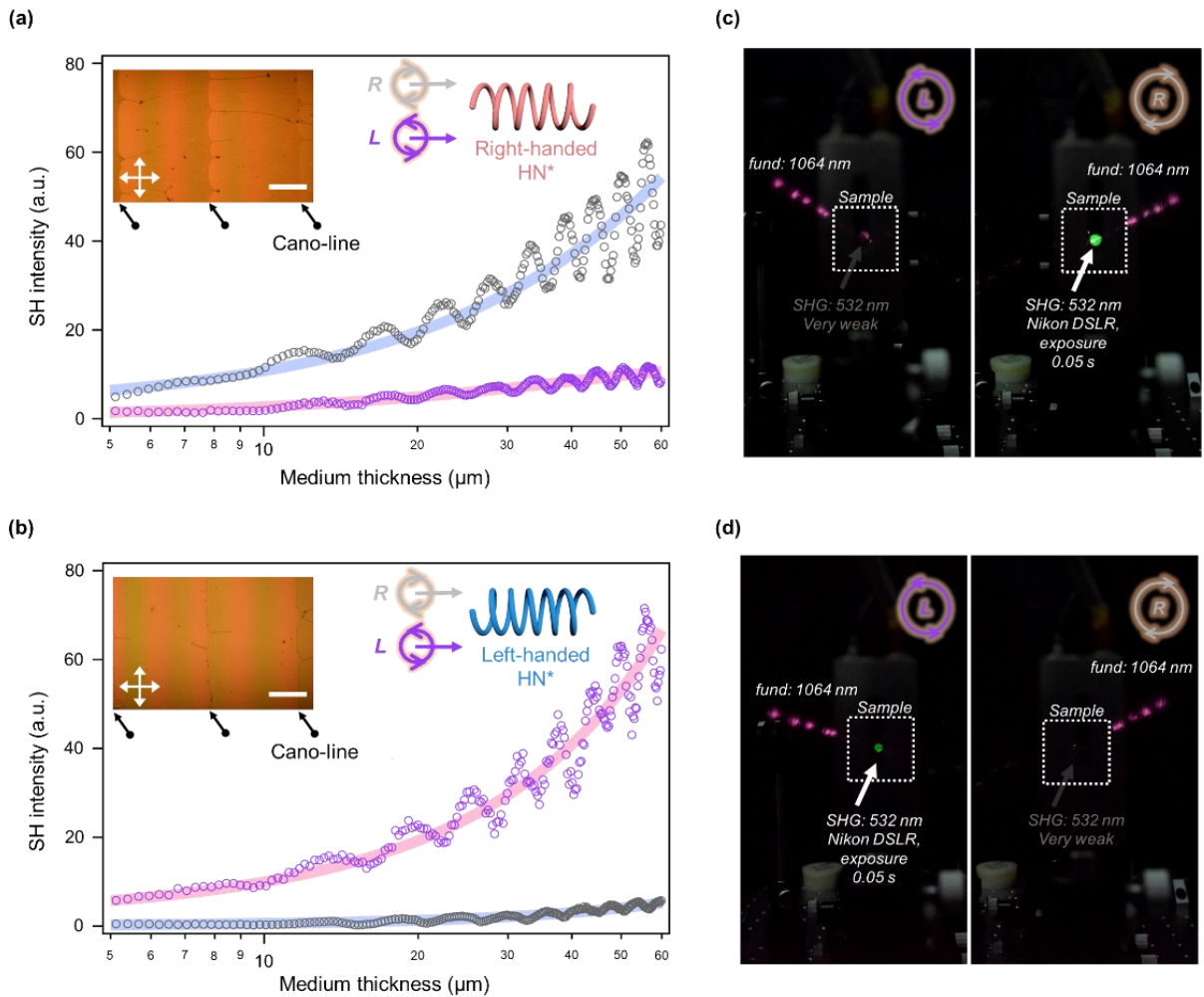

*Figure 3*. a) and b) the thickness dependence of the SH intensities of the right- and left-handed HN*, respectively, the horizontal coordinate is the log value. The pink and blue solid lines are the quadratic fittings of the SH signal. The SH intensity increases dramatically with thickness only when the handedness of the circular polarization of the fundamental wave is same to that of the HN* helix through the chiral HN* phase-matching pathway. The inserts show the texture of HN* in wedge cells, scale bar: 500 μm. c) and d) the visualization of SHG-CD for right- and left-handedness HN*, respectively, in 50-μm thick planar cells. The temperature for thickness dependencies of SHG and SHG-CD visualization was kept at 110 °C.

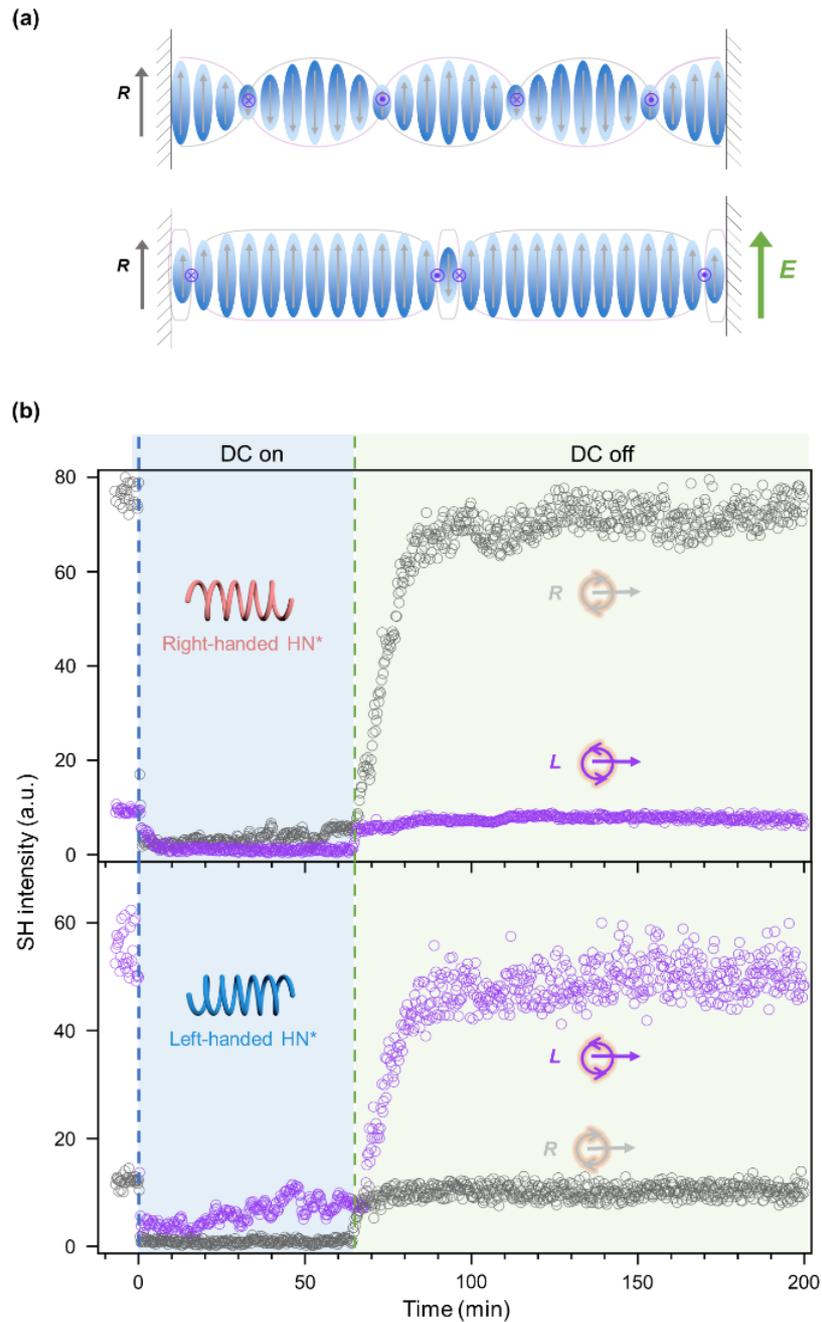

*Figure 4*. a) A schematic of a HN* structure without and under an in-plane electric field. The electric field is applied syn-parallel to the surface polarization direction. b) Time dependencies of SH intensity for left- (1.1%S811/RM734 (bottom)) and right-handed (1.1%R811/RM734 (top)) HN* under distinct circular polarizations. An interdigitated comb-shaped ITO cell with a thickness of 50 μm was used. Both the electrode width and gap between adjacent electrodes were 1 mm. A DC electric field of 1 mV/μm was applied from *t*=0 min, and the temperature was kept at 110 °C.

# Supporting Information

*High-g-Factor Phase-Matched Circular Dichroism of Second Harmonic Generation in Chiral Polar Liquids*

*Xiuhu Zhao[1], Jinxing Li[1], Mingjun Huang[1,2], Satoshi Aya\*,[1,2]*

[1]South China Advanced Institute for Soft Matter Science and Technology Advanced Institute for Soft Matter Science and Technology (AISMST), School of Emergent Soft Matter, South China University of Technology, Guangzhou, 510640, P. R. China

[2]Guangdong Provincial Key Laboratory of Functional and Intelligent Hybrid Materials and Devices, South China University of Technology, Guangzhou, 510640, P. R. China

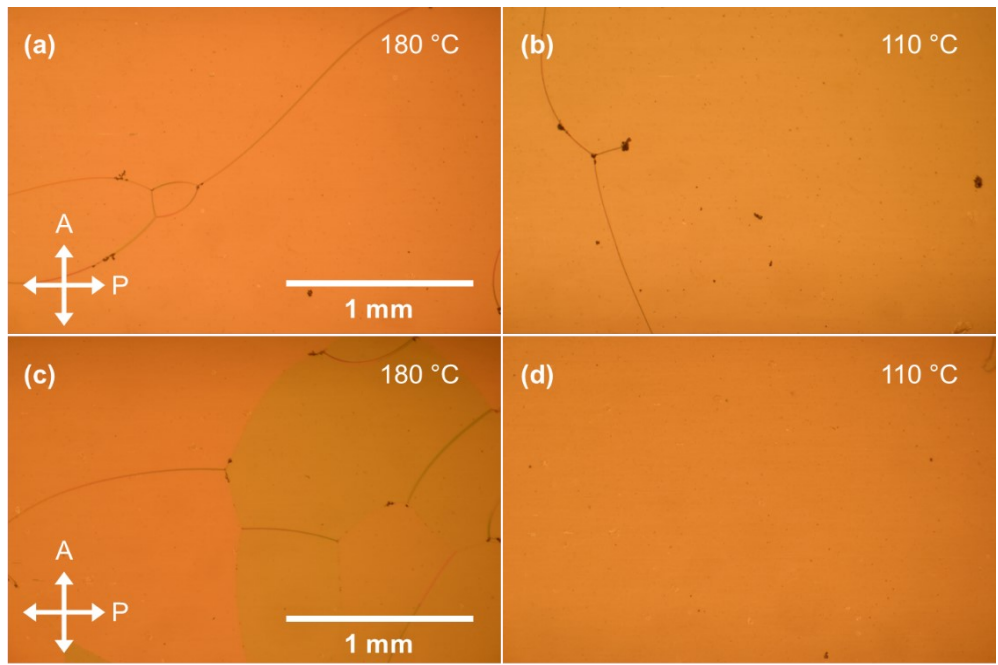

*Figure S1.* The PLM texture of two HN* in a 50 μm thickness cell. a) and b) the texture of 1.1%R811/RM734 at 180 °C and 110 °C, respectively; c) and d) the texture of 1.1%S811/RM734 at 180 °C and 110 °C, respectively. The scale bar is 1 mm.

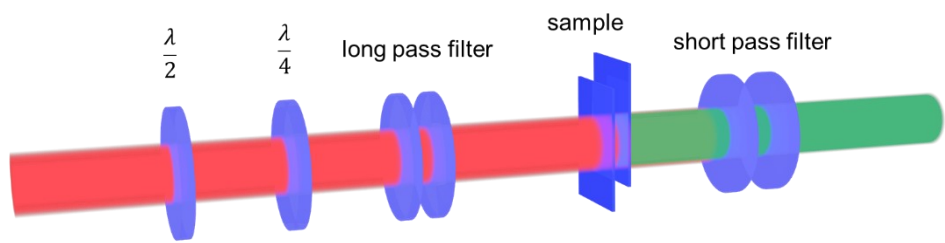

*Figure S2.* The optical path for SHG measurement. λ/2 and λ/4 represent the half-wave plate and quarter-wave plate, respectively.

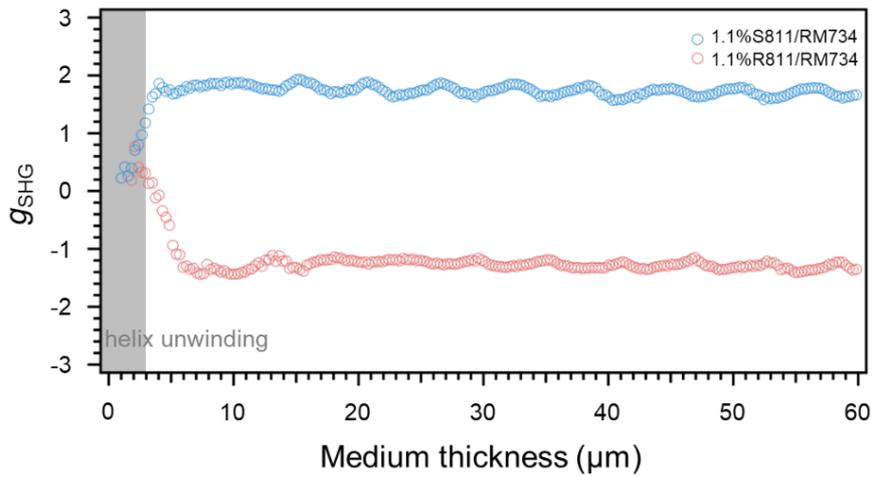

*Figure S3.* The thickness dependencies of the *g*-factor of right-handed HN* 1.1%R811/RM734 and left-handed HN* 1.1%S811/RM734, respectively. The gray part indicates the thinnest part of the wedge-shaped cell (0~3 µm), where the polarization field is in the unwinding state, and the SHG-CD effect is neglectable. Also, some defects are seen in this area, causing some fluctuations of *g*-factor.

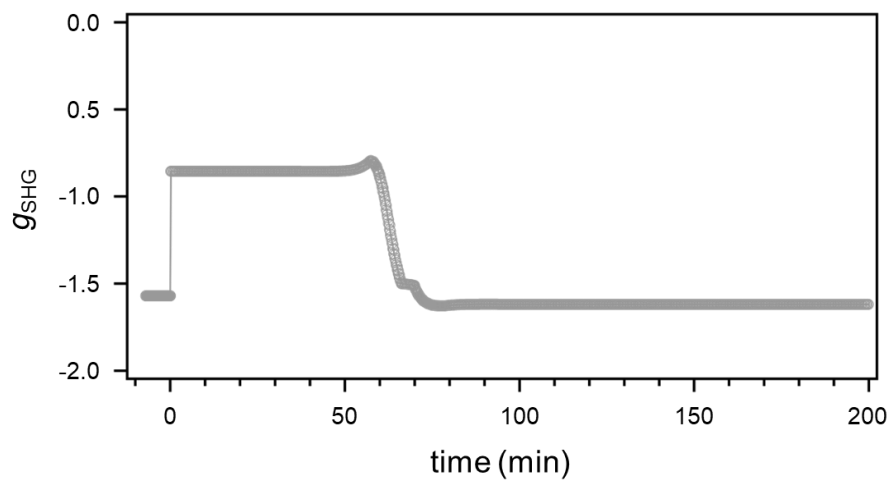

*Figure S4.* The *g*-factor of 1.1%R811/RM734 under a DC electric field of 1 mV/μm. The *g*-factor is calculated by first smoothing the SH signal under the electric field.

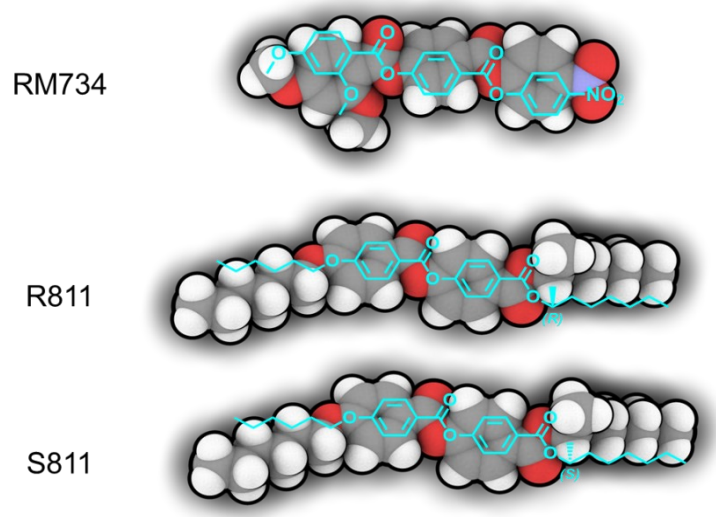

*Figure S5.* Molecular structure of N$_F$ liquid crystal RM734, chiral dopant R811 and S811